\documentclass[%
 reprint,
superscriptaddress,
 amsmath,amssymb,
 aps,
 twocolumn,
pra,
nofootinbib
]{revtex4-1}
\usepackage{pifont}
\usepackage[dvipsnames]{xcolor}
\usepackage[ruled,vlined]{algorithm2e}
\usepackage{svg}
\usepackage[utf8]{inputenc}
\usepackage[normalem]{ulem}
\usepackage{multirow}
\usepackage{enumitem}

\usepackage{graphicx}
\usepackage{dcolumn}
\usepackage{bm}
\usepackage{hyperref}
\hypersetup{colorlinks = true, linkcolor = [rgb]{0.19411,0.51882,0.667058}, urlcolor = [rgb]{0.125490,0.29542,0.1647058}, citecolor = [rgb]{0.75882,0.37411,0.14117}}
\usepackage{makecell}

\usepackage{setspace}
\usepackage{eurosym}
\usepackage{amssymb}

\usepackage{braket}
\usepackage[T1]{fontenc}

\graphicspath{{../images/}}
\usepackage[toc,page]{appendix}

\usepackage{braket}
\usepackage[bottom]{footmisc}
\def\>{\rangle}
\def\<{\langle}

\usepackage[bottom]{footmisc}
\SetKwInput{KwInput}{Input}                
\SetKwInput{KwOutput}{Output}              
\begin{document}

\title{
Unconditionally secure digital signatures implemented in an 8-user quantum network
}
\thanks{Correspondence and requests for materials should be addressed to
 Siddarth Koduru Joshi SK.Joshi@Bristol.ac.uk}

\author{Yoann Pelet}
\affiliation{Quantum Engineering Technology Labs, H. H. Wills Physics Laboratory \& Department of Electrical and Electronic Engineering, University of Bristol, Merchant Venturers Building, Woodland Road, Bristol BS8 1UB, United Kingdom. Now at: Université Côte d'Azur, CNRS, Institut    de Physique de Nice (INPHYNI), UMR 7010, Parc Valrose, 06108 Nice Cedex 2, France.}

\author{Ittoop Vergheese Puthoor}
\affiliation{Institute of Photonics and Quantum Sciences, Heriot-Watt University, United Kingdom}

\author{Natarajan Venkatachalam}
\affiliation{Quantum Engineering Technology Labs, H. H. Wills Physics Laboratory \& Department of Electrical and Electronic Engineering, University of Bristol, Merchant Venturers Building, Woodland Road, Bristol BS8 1UB, United Kingdom}

\author{S\"oren Wengerowsky}
\affiliation{ICFO-Institut de Ciencies Fotoniques, The Barcelona Institute of Science and Technology,08860 Castelldefels (Barcelona), Spain}
\author{Martin Lon\v{c}ari\'{c}}
\affiliation{Photonics and Quantum Optics Research Unit, Center of Excellence for Advanced Materials and Sensing Devices, Ru\dj{}er Bo\v{s}kovi\'{c} Institute, Zagreb, Croatia}
\author{Sebastian Philipp Neumann}
\affiliation{Institute for Quantum Optics and Quantum Information - Vienna (IQOQI) \& Vienna Center for Quantum Science and Technology (VCQ), Vienna, Austria}
\author{Bo Liu}
\affiliation{College of Advanced Interdisciplinary Studies, NUDT, Changsha, 410073, China}
\author{\v{Z}eljko Samec}
\affiliation{Photonics and Quantum Optics Research Unit, Center of Excellence for Advanced Materials and Sensing Devices, Ru\dj{}er Bo\v{s}kovi\'{c} Institute, Zagreb, Croatia}
\author{Mario Stip\v{c}evi\'{c}}
\affiliation{Photonics and Quantum Optics Research Unit, Center of Excellence for Advanced Materials and Sensing Devices, Ru\dj{}er Bo\v{s}kovi\'{c} Institute, Zagreb, Croatia}
\author{Rupert Ursin}
\affiliation{Institute for Quantum Optics and Quantum Information - Vienna (IQOQI) \& Vienna Center for Quantum Science and Technology (VCQ), Vienna, Austria}
\author{Erika Andersson}
\affiliation{Institute of Photonics and Quantum Sciences, Heriot-Watt University, United Kingdom}
\author{John G. Rarity}
\affiliation{Quantum Engineering Technology Labs, H. H. Wills Physics Laboratory \& Department of Electrical and Electronic Engineering, University of Bristol, Merchant Venturers Building, Woodland Road, Bristol BS8 1UB, United Kingdom}
\author{Djeylan Aktas}
\affiliation{Quantum Engineering Technology Labs, H. H. Wills Physics Laboratory \& Department of Electrical and Electronic Engineering, University of Bristol, Merchant Venturers Building, Woodland Road, Bristol BS8 1UB, United Kingdom}
\affiliation{Research Center for Quantum Information, Institute of Physics, Slovak Academy of Sciences, Dúbravská Cesta 9, 84511 Bratislava, Slovakia}
\author{Siddarth Koduru Joshi*}
\affiliation{Quantum Engineering Technology Labs, H. H. Wills Physics Laboratory \& Department of Electrical and Electronic Engineering, University of Bristol, Merchant Venturers Building, Woodland Road, Bristol BS8 1UB, United Kingdom}

\date{\today}

\begin{abstract}
The ability to know and verifiably demonstrate the origins of messages can often be as important as encrypting the message itself. Here we present an experimental demonstration of an unconditionally secure digital signature (USS) protocol implemented for the first time, to the best of our knowledge, on a fully connected quantum network without trusted nodes. Our USS protocol is secure against forging, repudiation and messages are transferrable. We show the feasibility of unconditionally secure signatures using only bi-partite entangled states distributed throughout the network and experimentally evaluate the performance of the protocol in real world scenarios with varying message lengths.
\end{abstract}

\maketitle


\section{\label{sec1}Introduction}
Digital signatures are cryptographic protocols that offer authenticity (i.e., the message originates from the correct user), integrity (i.e. the message cannot be altered) and non-repudiation (i.e the message is undeniably originates from the first sender and this can be verified even after being transferred by another user). Such protocols can replace handwritten signatures and allow centralised authorities to issue unforgeable IDs. 
They are also crucial to financial transactions, software distribution, and e-mails, where the sender has to exchange messages with many recipients, with the assurance that the messages cannot be forged or tampered with. Using digital signatures, a verifier does not need to contact the originator at the time of verification to prove the authenticity or integrity of the message.


Classical digital signature schemes such as RSA~\cite{Rivest1978} and ECDSA~\cite{Johnson2001} rely on public-key encryption, and provide only computational security. That is the security of such protocols is based on the assumed unproven hardness of inverting certain cryptographic functions. Essentially, this implies that the advancement of computer technologies or/and development of powerful algorithms could jeopardise the security of these protocols. 
This provides the main motivation for developing unconditionally secure signature (USS) schemes~\cite{Chaum1991,Pfitzmann1996,Hanaoka2000,Hanaoka2004,Swanson2011, Amiri2016} also referred to as information-theoretically secure signature schemes, as these schemes provide security which is not bounded by an adversary's computational power. Hence, USS schemes may be particularly useful in scenarios which require high security and where devices are limited in computational capability. For example, public key signature schemes need more computational power when compared to USS hash-based signature schemes. Because information theoretic security of any digital signature protocol requires secure shared randomness, scaling digital signature protocols up to many users is a major challenge. 

Chaum and Roijakkers~\cite{Chaum1991} constructed the first classical USS scheme that uses authenticated broadcast channels, encrypted authenticated channels and untraceable sending protocols. The disadvantage is that this scheme becomes less efficient when signing long messages. 
Pfitzmann and Waidner~\cite{Pfitzmann1996} developed a scheme that is built upon Chaum and Roijakkers, but allows a user to sign and verify long messages. However, this scheme still requires authenticated broadcast channels and untraceable sending protocols for implementation. Hanaoka \emph{et al.}~\cite{Hanaoka2000,Hanaoka2004} proposed a USS scheme that improves upon the previous classical USS schemes in terms of security, efficiency and by making the signature scheme re-usable, but requires additional trust assumptions. 

On the other hand, there are quantum versions of USS schemes that employ the principles of quantum mechanics to offer information-theoretic security. Gottesman and Chuang introduced the first quantum digital signature scheme~\cite{Gottesman2001} and Clarke \emph{et al.}~\cite{Clarke2012} achieved a first experimental realisation. Earlier schemes required long-term quantum memory but in later schemes~\cite{Andersson2006, Dunjko2014, Collins2014} this requirement was removed. By using quantum key distribution (QKD)~\cite{BB84} systems, one could also achieve more practical realizations of quantum USS schemes~\cite{Wallden2015, Amiri2016a, Donaldson2016}. Later, Puthoor \emph{et al.}~\cite{Puthoor2016} proposed a measurement-device-independent(MDI) scheme that is secure against detector side-channel attacks. With the advancement of MDI-QKD, this scheme has been experimentally realised~\cite{Yin2017, Roberts2017} for networks with three participants. 


In this work, we provide the first experimental implementation of a USS scheme~\cite{Amiri2016} on a quantum network comprising more than $3$ participants. USS schemes use secret keys rather than public keys, and the secret keys are distributed among participants before messages can be signed or verified. 
Our network consists of a physical layer, a quantum layer where bi-partite entanglement is shared and a communication layer where classical data is exchanged.
Further, our network is fully connected in the quantum and communication layer which means that every one of the 8 users directly generates a secure QKD key with every other user.
Our USS protocol is implemented in the communication layer of the network after the QKD keys have been generated.
  This is also the first, to the best of our knowledge, demonstration of transferable information-theoretically secure digital signatures in a large network. 
The transferable nature of our digital signatures ensures that a message accepted by a recipient, will also be accepted by another recipient if the message is forwarded. {This implementation has potential applications in voting, secret sharing and in forming central certification authorities for the future quantum internet.} 

\section{Quantum network testbed}
We use an eight-user entanglement-based fully connected quantum network~\cite{joshi2020trusted} as a testbed to implement the USS scheme. The network generates bi-partite polarisation entanglement in a central source which is multiplexed and distributed among eight users such that all pairs of users share entanglement. A single optical fibre connects each user to this central source and multiplexing unit. The users themselves consist of a polarisation analysis setup (with a passive basis choice~ \cite{GISIN1999_Switch}), 2 single photon detectors and a time-tagging unit connected to a computer. 

\begin{figure*}[htb]
	\begin{center}
   \includegraphics[width=2\columnwidth]{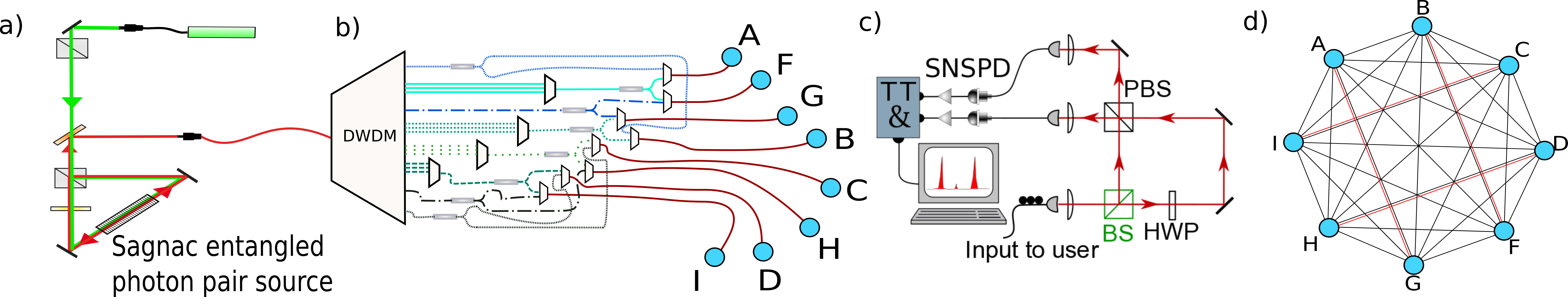}
	\caption{Elements of the experimental setup. (a) Sagnac source to generate polarisation entangled photon pairs. (b) Frequency demultiplexing and allocation of the channels to each users. (c) Polarisation analysis setup provided to every user. (d) Topology of the resulting network. The red lines represent the premium links that are faster than the regular ones.}
	\label{topology}
    \end{center}
\end{figure*}

In our experiment, the network was operational for about 18 hours and 24 minutes.
Each pair of users processes its measurement results following the BBM92~\cite{bbm92} protocol using a security parameter of $10^{-5}$ and accounting for finite-size key effects~\cite{FiniteKey_ScaraniRenner2008} to simultaneously generate 32 secure keys across the network. This forms a fully connected topology as shown in fig~\ref{topology} with 28 regular links and 4 premium links corresponding to users provided with 2 wavelength channels instead of 1 for other regular users.
We implemented a relatively fast but inefficient error correction based on Low-Density Parity Check and privacy amplification to rapidly generate secure keys from each 20 minutes block of collected data. Every user maintains 7 secure key stores, one for communication with every user, which are updated every 20 minutes. The USS protocol then uses keys from these key stores.

\section{\label{sec2}Unconditionally Secure Signatures}
In this section, we provide a brief description of the USS protocol we implement~\cite{Amiri2016}. We consider a protocol with $8$ participants: a sender (Alice denoted by 'A' in Figure.~\ref{topology}) and $7$ receivers (B,C,D,F,G,H,I). 
Here, the sender uses the shared secret keys to give independently to each recipient a number of hash functions (i.e. $\epsilon-$Almost Strong Universal (ASU) hash functions~\cite{Carter1979}) which will be used to authenticate a message by appending tags (hash values) to the message being sent. To verify the signature, the receivers will apply their hash functions to the message, and record the number of mismatches between their hash values and the appended tags. The receiver will only accept the message if he/she finds less than a threshold amount of mismatches. 

In order to ensure that the message is transferable (i.e. the message can be transferred between receivers while keeping the signature from the original sender), each recipient will group the hash functions received from the sender into $7$ equally sized sets (of size $k$). If the sender knew which participants will use which hash functions to check the authenticity of a message, then a malicious sender could ensure that one recipient accepts the message while another does not. This could be achieved by choosing the appropriate tags to be appended and thereby breaking the transferability of the protocol. To ensure that this does not happen, the recipient then sends one of the seven sets to every other recipient by using encrypted channels that are constructed using QKD, and also keeps one for himself. 
The recipients then independently uses the hash functions from each of the $7$ sets to verify the received signatures (as explained in Section~\ref{sec:methods}). Therefore, to ensure transferability, we have different levels of verification, and a limit on the number of times a message is guaranteed to be transferable in sequence. Also, the number of dishonest participants in the protocol would have an impact on the verification levels. Trivially, the protocol cannot be made secure if more than half the number of participants are dishonest. Hence, as described in~\cite{Amiri2016}, we set the maximum fraction of dishonest recipients ($d_R$) to be involved in the protocol as $d_R = l_{max}/N$ with $l_{max}$ being the maximum level of transferability a given network can achieve. $l_{max}=1$ for our seven users network, and N the number of recipients. This number of dishonest participants can be increased if we choose to decrease $l_{max}$.

\section{Methods}
\label{sec:methods}
\begin{figure}
   \includegraphics[width=0.95\columnwidth]{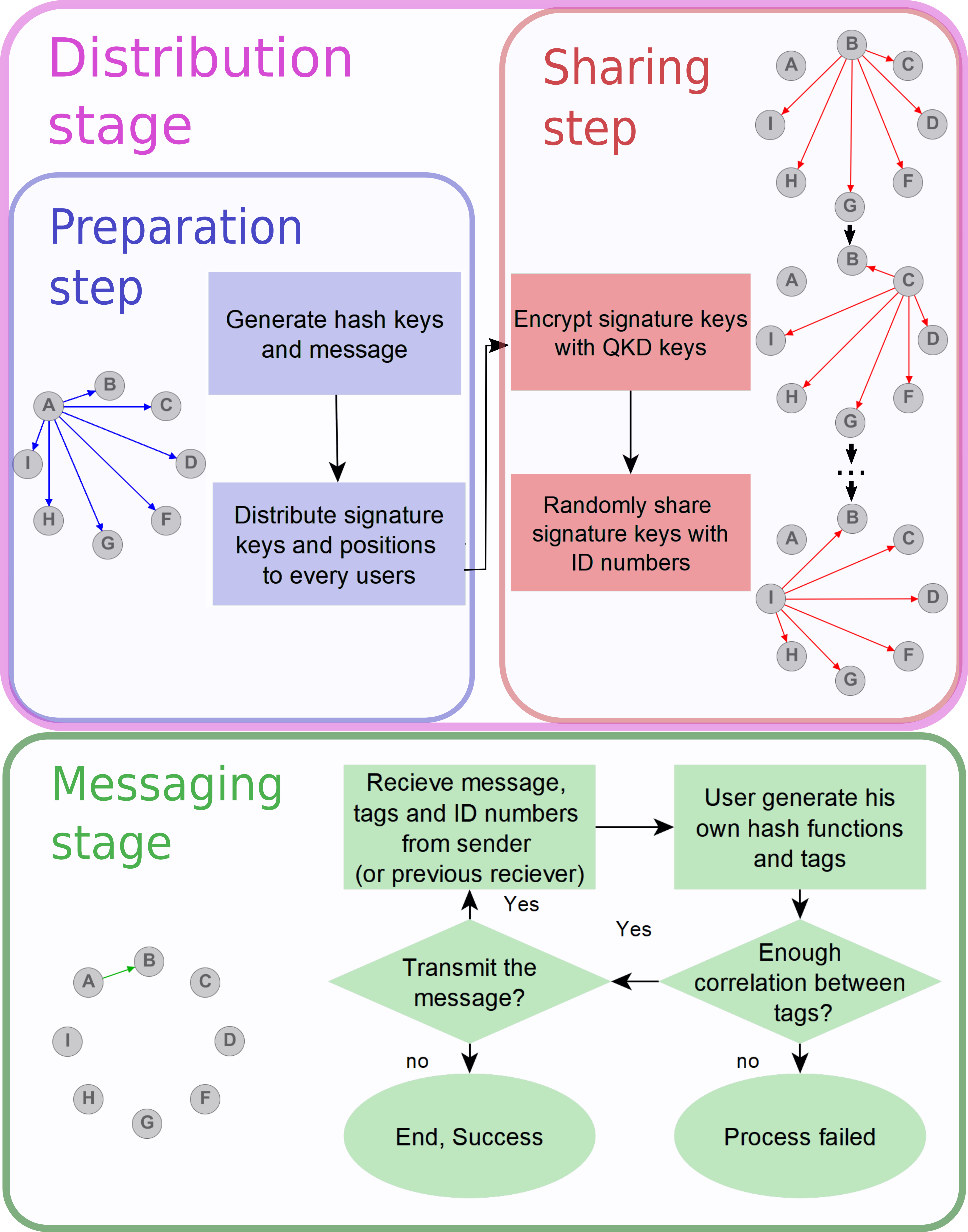}
	\caption{Flowchart of the stages and steps involved in the USS protocol. The interaction of the participants in the protocol are indicated by the arrows. Blue arrows represent the preparation step in the protocol, red arrows represent the sharing step, and the green for the messaging stage.}
	\label{scheme}
\end{figure}

As in the previous signature protocols, there are two stages - the distribution stage and the messaging stage. The workflow of these stages in the signature protocol is shown in Figure.~\ref{scheme}.

\textbf{Distribution stage:} This stage primarily comprises of two steps:

1. We call our first step the $preparation$, where the sender generates a set of $N^2k$ keys which are referred to as the $signature$ $keys$. Here, $k$ is a security parameter which is a significant factor in determining the security level and transferability of the signature scheme. The $signature$ $keys$ are bit strings which are extracted from the QKD key store. Each recipient receives $Nk$ keys from the sender. Those keys are directly taken from the secret key stores which are maintained via our QKD network.

2. After the $preparation$, the next step is $sharing$. Each recipient uses his encrypted channels (secured by QKD keys) to randomly exchange $k$ keys and their respective key identification number from his original set to every other receivers.
We note that in a practical implementation of the protocol each signature key needs an ID number to avoid confusion. 
 
 For our network of $8$ participants ($1$ signer and $7$ recipients) network, each of the recipients will send $k$ keys to every other recipients. Following this sharing process, each recipient has $7k$ keys and their ID numbers.
 In the future, the participants use these to check the validity of a signed message. We note that once QKD keys have been used to encrypt data shared via secure channels these keys cannot be reused.

\textbf{Messaging stage:} 

1. To send a message $m$, the sender (for example Alice) generates a list of $N^2k$ tags which constitutes the signature for the message. These tags are generated by a Python algorithm which initially transforms the $signature$ $keys$ to $signature$ $hash$ $functions$ that are later applied to the message $m$. The message along with the signature is sent to a recipient (for example Bob).

2. For a fixed level of verification ($l$), Bob, the recipient performs a test as described in~\cite{Amiri2016} for the message $m$ and the signature (list of $N^2k$ tags). First, Bob uses his signature keys to generate the hash functions. He then uses these hash functions on the message and obtains a list of tags ($7*k$ in our example). Also having the key positions, he can then compare the tags that he generated with the tags that are sent by Alice. Importantly, the test checks whether Alice's signature matches with what Bob is expected to obtain for the level $l$, but would allow a certain number of mismatches. The fraction of the number of mismatches allowed for a given transferability level $l$ is $s_l$ with $s_-1$). For each verification level, all participants will perform $7$ such tests. In our network, $l_{max}$ is 1 and for a security parameter of $10^{-10}$, we set the parameters $s_{1}, s_{0}, s_{-1}$ as defined in~\cite{Amiri2016} to be $s_{1} = 0.005$, $s_{0} = 0.252$ and $s_{-1} = 0.499$. These parameters need to be evenly spaced between 0 and 0.5 to ensure equal failure probabilities for all levels of transferability. So the best theoretical choice for $s_{l_{max}}$ ($s_{1}$ for our 7 users implementation) and $s_{l_{min}} = s_{-1}$ are $0+\epsilon_1$ and $0.5-\epsilon_2$, where $\epsilon_{1,2}$ are arbitrarily small positive numbers. 
For our experimental implementation we chose the minimum fractional error rate we will tolerate to be 0.005. Thus we are robust against up to 30 erroneous hash functions (in our experimental implementation) from uncorrected classical communication without affecting the protocol.  For the maximum bound, we experimentally set $s_{-1} = 0.499$ such that we can safely stop the protocol with at least 6 erroneous hash functions (since we have 7 receivers) before the theoretical limit. Such considerations are useful if there is an experimental or statistical error in estimating the fraction of erroneous hash functions.

3. Bob will accept the signature as valid at level $l$ if the number of tests passed is greater than a threshold ($\delta_l$), which is given in ~\cite{Amiri2016} as $\delta_l = 0.5 + (l + 1)d_R$. Each recipient can accept or reject a message without interacting with the other participants in the messaging stage. It is important to note that the parameter $\delta_l$ depends on factors such as the fraction of dishonest participants that the protocol can tolerate, the desired transferability level of the protocol and the expected errors in the secret bit strings that are generated using QKD. Also, the maximum transferability level for a given fraction of dishonest participants is set by the condition provided in~\cite{Amiri2016} which is $(l_{max} + 1)d_R < 1/2$.

4. To forward the message, Bob sends the message $m$ along with the signature to the next recipient. The new recipient then proceeds with the messaging stage from step 2.

\subsection{\label{sec4}Security analysis}

We follow the security definitions provided in~\cite{Amiri2016} and show that the protocol is proven to be secure against forging, repudiation and that messages are transferable. Forging means that a dishonest participant (say Eve), who is not the signer, must output a message-signature pair that will be accepted by a participant (say Charlie) at a transferability level $l$. Equation $10$ in~\cite{Amiri2016} shows the probability of forging as
\begin{equation}
    P(\text{Forge})\approx N^{2}(1-d_{R})^{2}p_t.
    \label{forging}
\end{equation}
Here, $N$ is the number of participants (excluding the signer) and $d_{R}$ is the fraction of dishonest participants of the protocol. The parameter $p_t$ is the probability that the dishonest participant can create a message-signature pair that will pass the test performed by Charlie for the keys or functions received from the honest participants of the protocol.

In order to break the transferability of the protocol, a dishonest sender, Alice, must generate a signature that would be accepted by a honest participant (say Bob) at transferability level $l$ but would be refused by another honest participant (say Charlie) at transferability level $l' < l$. The probability to generate such a signature is given in~\cite{Amiri2016} as Equation $16$
\begin{equation}
    P(\text{Non-Transferability})\approx N_{p}(N(\delta_{l}-d_{R})+1)p_{m_{l,l-1}}
    \label{non_transferability}
\end{equation}
where $p_{m_{l,l-1}} \leq \exp\Big(-\frac{(s_{l-1}-s_{l})}{2} k\Big)$.
Here, $N_{p}=[N(1-d_{R})][N(1-d_{R})-1]/2$. Repudiation is a special case of non-transferability. Hence, security against repudiation can be obtained by substituting $l=0$ (lowest level of transferability) in equation~\eqref{non_transferability} and this gives us the probability of repudiation $P_r$. 

Using equation \eqref{non_transferability}, we set the probability of repudiation to be $10^{-10}$ and find the optimal value for the security parameter $k$ (906 for a 7 user network with $l_{max}=1$). The probability of failure of the protocol is the maximum of all the probabilities which includes forging and repudiation. Our numerical study shows that $P(\text{Forge})$ is very low and is negligible when compared to $P_r$. Therefore, we can consider $P_r$ to be the failure probability of the signature protocol.
 
 For our chosen value of $P_r = 10^{-10}$ Table~\ref{tab:scale} shows the number of secret bits needed to sign different message lengths and number of users. This table also compares the theory to our experimental implementation. We note that while the theoretical work predicts a $N^2k\log a$ term our implementation uses $N^2ka$. This is because the theoretical work is in the limit of large $N$. Our implementation uses a hash key the same length as the message to ensure security of the protocol for small $N$. This can be done without any loss in security but it does require more data to be transmitted.  
 
 \begin{table*}[ht]
\centering
\begin{tabular}[t]{lcc}
\hline
~&Theoretical work&This implementation\\
\hline
Key required&$O\big(N^{2}k(\log\,a + \log\,Nk)\big) $&$N^{2}ka+N(N-1)(a + \log\,kN)$\\
Secret bits required to sign 1 bit & 16515 & 14958 \\
Secret bit required to sign 8 bits & 21888 & 35898 \\
\hline
\end{tabular}
\caption{Comparison between theoretical protocol and our implementation for 7 users.\label{tab:scale}}
\end{table*}%

 \subsection{The quantum network testbed}

The source is pumped by a continuous-wave 775.1\,nm pump beam which undergoes broadband ($\approx 60$\,nm Full Width at Half Maximum) type 0 spontaneous parametric downconversion in a 40\,mm long MgO:PPLN (Magnesium Oxide doped Periodically Poled Lithium Niobate) crystal in a Sagnac configuration. This generates broadband entangled photons with a central wavelength of 1550.2\,nm in the Bell state $\ket{\phi^+} = \frac{1}{\sqrt{2}} \left( \ket{HH} + \ket{VV} \right)$, where H (V) represents a horizontally (vertically) polarised photon. We use dense wavelength division multiplexing (DWDM) in a 100\,GHz spaced International Telecommunication Union standard to split these entangled photons into 16 channels spaced symmetrically about the centre wavelength of 1550.2\,nm. Due to energy conservation during the downconversion process only symmetrically located channels carry entangled photon pairs.

Each of the 8 users consists of a beam splitter (BS) with a half-wave plate in one output arm to choose the measurement basis as either H/V or diagonal/antidiagonal. A polarising beam splitter (PBS) is used for the polarisation measurement and the two output ports are coupled into two single photon detectors. A path-length difference of 3.7 ns between the two output arms of the BS allows us to use the same PBS and two detectors for all measurement bases and still distinguish the measurement outcomes.

\section{Results}
$\textbf{Network topology.}$ The topology of the USS protocol network is shown in Figure.~\ref{topology}. The USS protocol with $8$ participants is implemented in an entanglement-based quantum network. We believe that this is the first implementation of a USS protocol in a large quantum network with previous implementations limited to only $3$ participants. All participants are completely connected to each other with $4$ additional premium links connecting $4$ different pairs of users.  We assume participant `A' in our network to be the signer and the others to be the recipients. The goal of a signature scheme is to demonstrate the authenticity of a signed message to multiple recipients and we show that this is possible for a given signed message even if it is transmitted through the whole network. In addition, the signature scheme is secure against the threats of forging, repudiation and non-transferability. 

\begin{figure}[htb]
	\begin{center}
   \includegraphics[width=0.9\columnwidth]{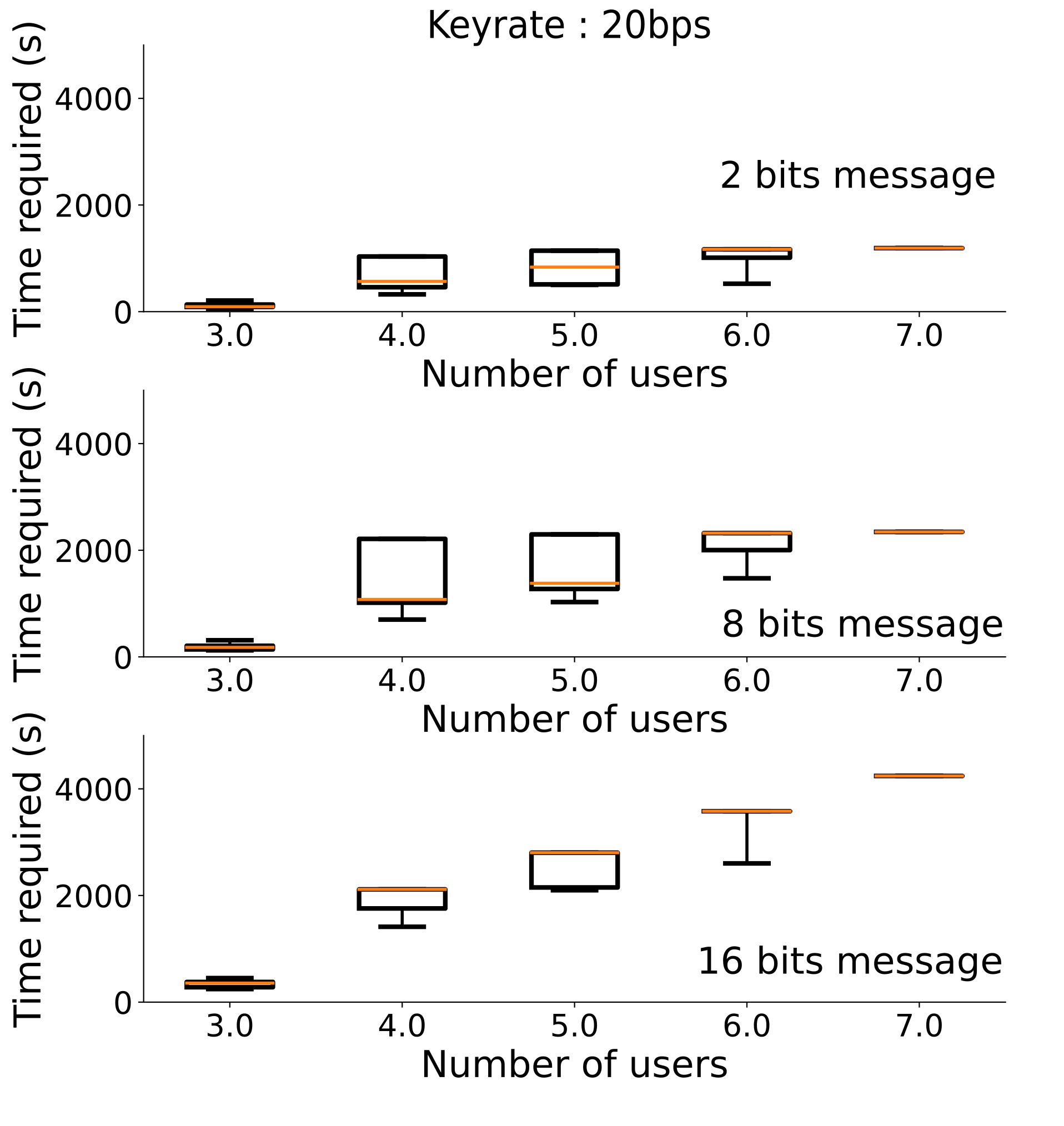}
	\caption{Experimental time required to generate enough keys for the distribution stage (with an optimised value $l_{max}=1$) as a function of the number of users for messages of lengths 2, 8 and 16 bits. Statistics are built by taking all possible combinations of users for a given number. The orange bar is the median, the box defines the interquartile range and the whiskers are the extreme values. The keyrate is for the worst link in the network since it will limit the overall speed of the process in the network.}
	\label{time(msg_length)}
    \end{center}
\end{figure}

$\textbf{Generation of signature keys.}$ For the signature protocol, we require data transfer (signature keys) between all participants and therefore this consumes more quantum keys than a normal encryption protocol. Figure~\ref{time(msg_length)} shows the time required to generate enough {QKD keys} on our network test-bed to set up a signature scheme sufficient for signing different numbers of message bits. The testbed is a fully connected network, however, practical use cases can include scenarios where only a subset of $N$ recipients participate in our USS protocol and the other users continue to generate QKD keys or participate in other protocols. We note that in terms of network performance this is the worst-case scenario as key generation rates are significantly improved if the number of connections is limited. The performance of any individual link in our network is limited by the detectors in use and since we multiplex signals from many links on to the same detector, having all links active is the worst-case scenario.
We compute the performance statistics for different numbers of participants in our USS protocol. For example, 4 users means $1$ signer and $N = 3$ recipients. 
From our experimental statistics that, we find that the size of the message has an impact on the time required to generate enough QKD keys. 

\begin{figure}[htb]
	\begin{center}
   \includegraphics[width=0.95\columnwidth]{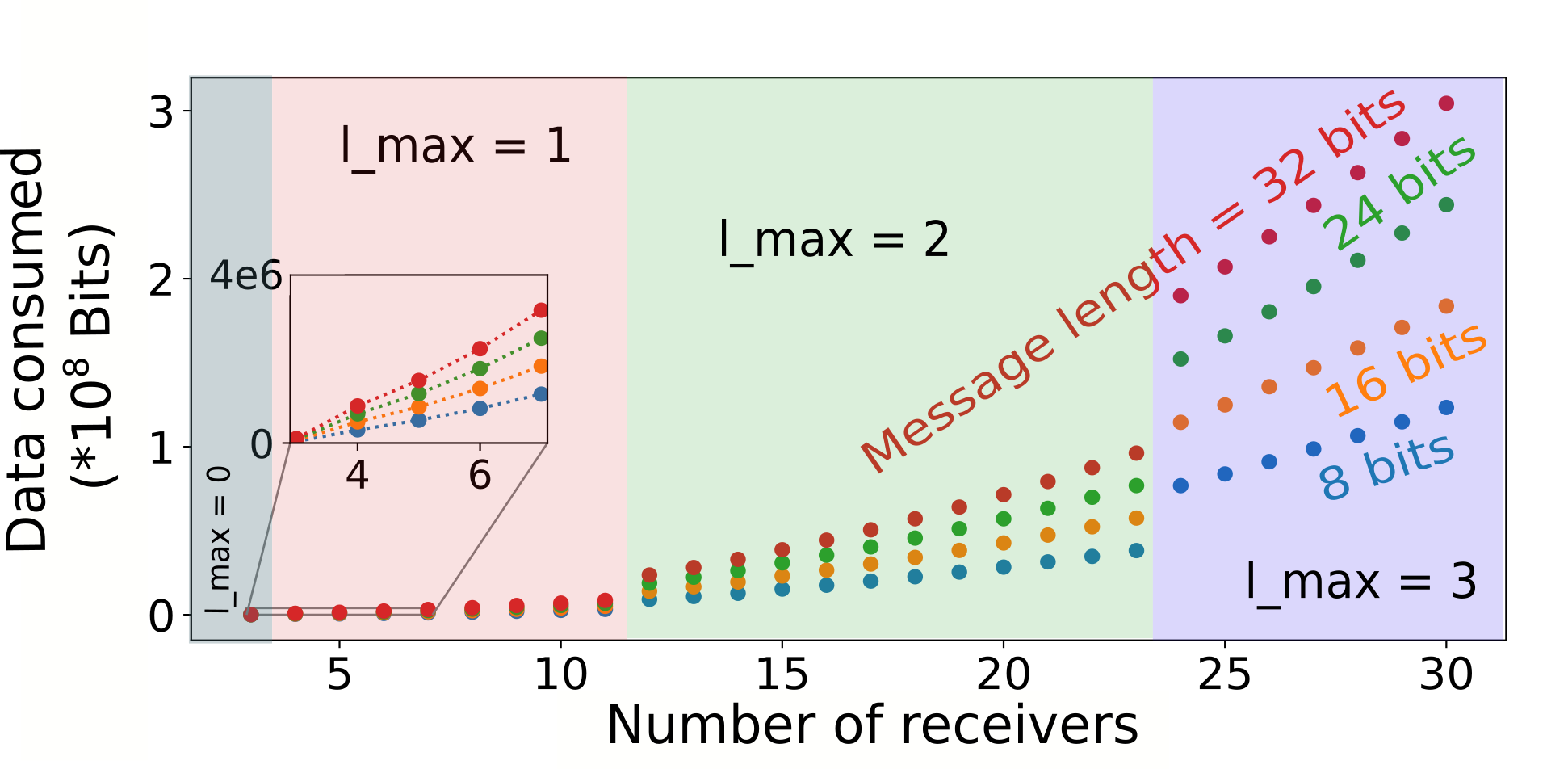}
	\caption{Data consumed during the distribution stage as a function of the number of potential recievers in the network. For each message length and considering the maximum level of transferability, the amount of data required is calculated.}
	\label{data_per_users}
    \end{center}
\end{figure}

$\textbf{Data consumption.}$ From Figure~\ref{data_per_users}, we find that the amount of data consumed increases with the number of recipients and with the levels of transferability that is desired for the protocol. For the size of the current network, we find that it is not relevant to use more than 2 levels of transferability since $l_{max}=1$. Hence, the data consumed is far lower when compared to bigger networks. 
This is illustrated by the difference in the 4 regions of the graph shown in Fig.~\ref{data_per_users} corresponding to 1, 2, 3 or 4 levels of transferability.
 We have performed a simulation of the signature protocol for bigger networks and calculated the data consumption with respect to the number of receivers and message size. The results are depicted in Figure~\ref{data_per_users}. There are different factors that control the signature verification, such as the number of dishonest participants and the number of transferability levels. We can see that the number of transferability levels is lowered when the number of dishonest recipients increases. By using $(l_{max} + 1)d_R < 1/2$, we can obtain a maximum level of transferability $l_{max}$. 
 We note that the first level of transferability is $l_{max} = 0$. 
 Figure \ref{data_per_users} shows the data consumption in four different coloured zones, and the plot is produced for the worst-case scenario. Each colour represents the maximum transferability level of the protocol. For example, in the last zone (coloured blue), if we set the protocol to have more than $3$ dishonest receivers then we get a maximum transferability level of $3$ instead of $4$.  It is important to note that for a given $l_{max}$ the variation of data consumption with respect to the number of recipients is close to linear, and the scenario doesn't improve when there are more levels of transferability. Hence, it is important to set appropriate $l_{max}$ depending on the application of the signature protocol.

\begin{figure}[htb]
	\begin{center}
   \includegraphics[width=0.95\columnwidth]{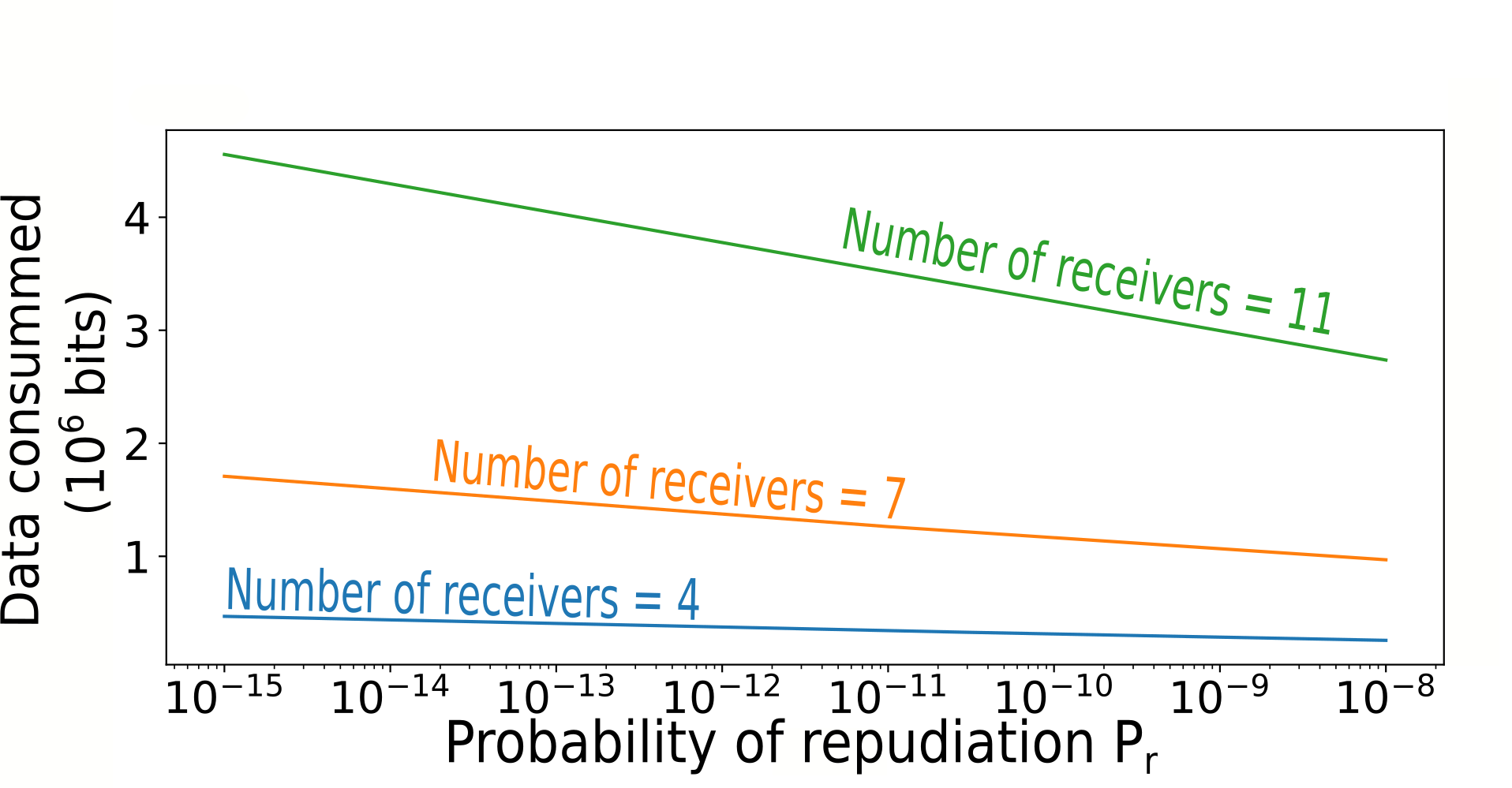}
	\caption{Data consumed (bits) as a function of $P_{r}$ (log scale) for different numbers of receivers for 8 bits messages.}
	\label{data_pr}
    \end{center}
\end{figure}

$\textbf{Implementation in a large network.}$ As mentioned in the previous section, the failure probability of the protocol is mainly considered as the probability of repudiation since the forging probability is very low. We show in Figure \ref{data_pr} the variation of data consumption for different numbers of participants with regard to $P_r$. We note that the data consumed also increases with the number of participants. 
It may be possible to optimise amount of secure key consumed for each use case.
In our scenario, considering $8$ participants and for $P_r = 10^{-10}$, we get the data consumption to be around $5$ Mbits.

\begin{table*}[ht]
\centering
\begin{tabular}[t]{lccccccc}
\hline
~&USS (our implementation)&\cite{Donaldson2016}&\cite{Yin2017}&\cite{Collins2017}& \cite{Roberts2017}& \cite{yin2017satellite}&\cite{Lu2021}\\
\hline
repetition rate&continuous&100MHz&75MHz&1GHz&1GHz&75MHz&1GHz\\
Number of receivers &7&3&3&3&3&3&3\\
Security parameter&$10^{-10}$&$10^{-4}$&$10^{-9}$&$10^{-10}$&$10^{-10}$&$10^{-10}$&$10^{-9}$\\
Signature time per bit&\ \ 1041s (73s for 3 receivers)&$10^{6}$s&66840s&27s&45s&40hrs&14,9s\\
\hline
\end{tabular}
\caption{Comparative table for different QDS implementations. \label{tab:QDS}}
\end{table*}%

\section{\label{sec6}Conclusion}

This work demonstrates the experimental implementation of a USS protocol in a QKD network with one sender and up to $7$ recipients with probabilities of forging and repudiation below $10^{-10}$. With the help of QKD links in our quantum network, we are able to generate information-theoretically secure digital signatures for the network users. This enables an easy transition to real-world applications. A comparison to other state-of-the-art implementations of QDS schemes is shown in Table~\ref{tab:QDS}.
To the best of our knowledge this is the first experimental implementation of an unconditionally secure digital signature scheme on a many user quantum network. Further improvements in the QKD key consumption efficiency of such protocols are necessary for their widespread adoption in large networks.

To compare our protocol with other implementations of quantum digital signatures, we look at the least amount of secure data distributed throughout the network for a given number of users (and given probability of failure) in order to transmit the required hash functions.
To the best of our knowledge, this implementation scales better with the number of users than any previously reported quantum digital signatures. 
Which makes it the ideal candidate for larger networks. 
We illustrate this point with the fact that our results are faster than several previous realisations with just 3 users (see Table~\ref{tab:QDS}).

The USS scheme uses QKD keys to extend the functionality of the quantum network. Together with a variety of other protocols such as authentication transfer~\cite{solomons2021scalable}, network flooding~\cite{solomons2021scalable} and a suite of anonymity protocols~\cite{huang2020experimental}, our USS protocol enables a wide range of end user application on quantum networks.

The additional security guarantees offered by the USS schemes over public-key digital signatures is significant for their use in many applications such as forming a certification authority, secret sharing or voting. 
In real world networks that combine slower direct links (e.g., entanglement distribution via satellite~\cite{Yin2017}, long fibres~\cite{wengerowsky2020passively,chen2020sending} or eventually via repeaters) with faster links based on trusted-nodes, signatures could be used to verify the original sender or even securely estimate the network topology without using a central authority. Also, these schemes play a crucial role in the future of the global digital economy where information security is essential. 

\subsection*{Acknowledgements}
The research leading to this work has received funding from United Kingdom Research and Innovation's (UKRI) Engineering and Physical Science Research Council (EPSRC) Quantum Communications Hub (Grant Nos. EP/M013472/1, EP/T001011/1) and equipment procured by the QuPIC project (EP/N015126/1). We also acknowledge the Ministry of Science and Education (MSE) of Croatia, contract No. KK.01.1.1.01.0001. We acknowledge financial support from the Austrian Research Promotion Agency (FFG) project ASAP12-85 and project SatNetQ 854022.  We would like to thank Thomas Scheidl for help with the software used to run the original experiment and Mohsen Razavi \& Guillermo Curr\'as Lorenzo for their help proving the security of the implementation of the original network experiment as well as Sebastien Tanzilli and Anthony Martin for helpfull discussions.


\subsection*{Author Contributions}
YP, SKJ, IVP and DA implemented the signatures protocol on the quantum network. SKJ, DA, SW, ML, SPN, BL, ZS, MS, JGR and RU built the original network. NV with the help of BL wrote software to process the quantum secured keys from the measurements of each user in the network. The project was conceived by SKJ and EA. 
SKJ, EA and JGR supervised the project and contributed to the ideas. SKJ was the team leader. The paper was written by YP, DA, IVP and SKJ and all authors read and contributed to improving it.

\bibliographystyle{plain}

\bibliography{ref}

\appendix






\section{Error tolerance}
\begin{figure}[htb]
	\begin{center}
   \includegraphics[width=0.95\columnwidth]{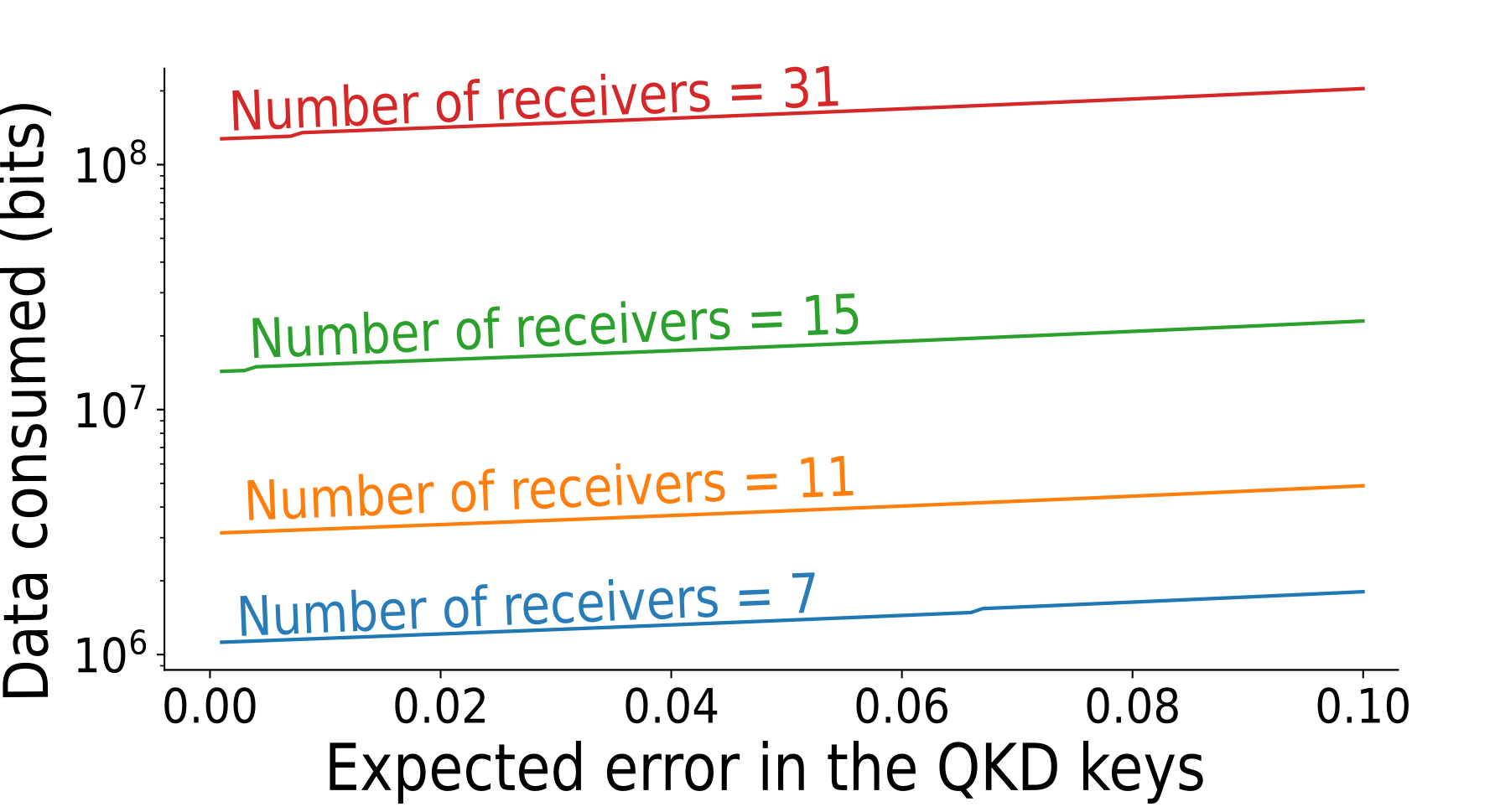}
	\caption{Data consumed during the distribution stage as a function of the estimated error in the quantum keys for different numbers of receivers. The amount of data required is calculated to have a probability of repudiation of $10^{-10}$.}
	\label{data_error}
    \end{center}
\end{figure}

It is of interest to note that our digital signature protocol can continue to function despite errors in the QKD keys (i.e. a mismatch in the bit strings) between two users. The small modifications necessary to the protocol are described later on in this section. To demonstrate this error tolerance we randomly simulated errors in our QKD keys. The resulting increase in the QKD key consumption rate is shown in Fig \ref{data_error}.

It may be possible to to exploit the above error tolerance of the USS protocol to increase the overall rate at which signatures can be generated without altering any experimental parameters. 
We generate quantum keys using experimental data from QKD links within the network test-bed. Often, some imperfections are unavoidable and error correction must be applied before privacy amplification.
We use an LDPC-based error correction algorithm where information is exchanged back and forth iteratively between users until all errors in the sifted keys are corrected. All error correction information exchanged is available to a malicious party and privacy amplification must be used to ensure security. By performing fewer error correction iterations, it is possible to obtain a sifted key and consequently a secure key with a small error rate. Since less information was leaked, this secure quantum key can be longer.
We note that all errors in the sifted key contribute to the QBER and the amount of privacy amplification needed because of a certain QBER value remains the same as in the BBM92 protocol. Instead we only reduce the amount of information leaked during error correction.  

The increase in length of the QKD key (by reducing the amount of error correction performed) could possibly outweigh the excess key consumed for error tolerance in our protocol.


By changing the verification condition for the highest transferability level from $100\%$ correlations to $100\%-expected\ errors$, we get a protocol that is resistant to a certain degree of errors in the data. Changing this verification condition this way has an impact on the data consumed for the protocol as showed in Fig \ref{data_error}. 

\begin{figure}[htb]
	\begin{center}
   \includegraphics[width=0.95\columnwidth]{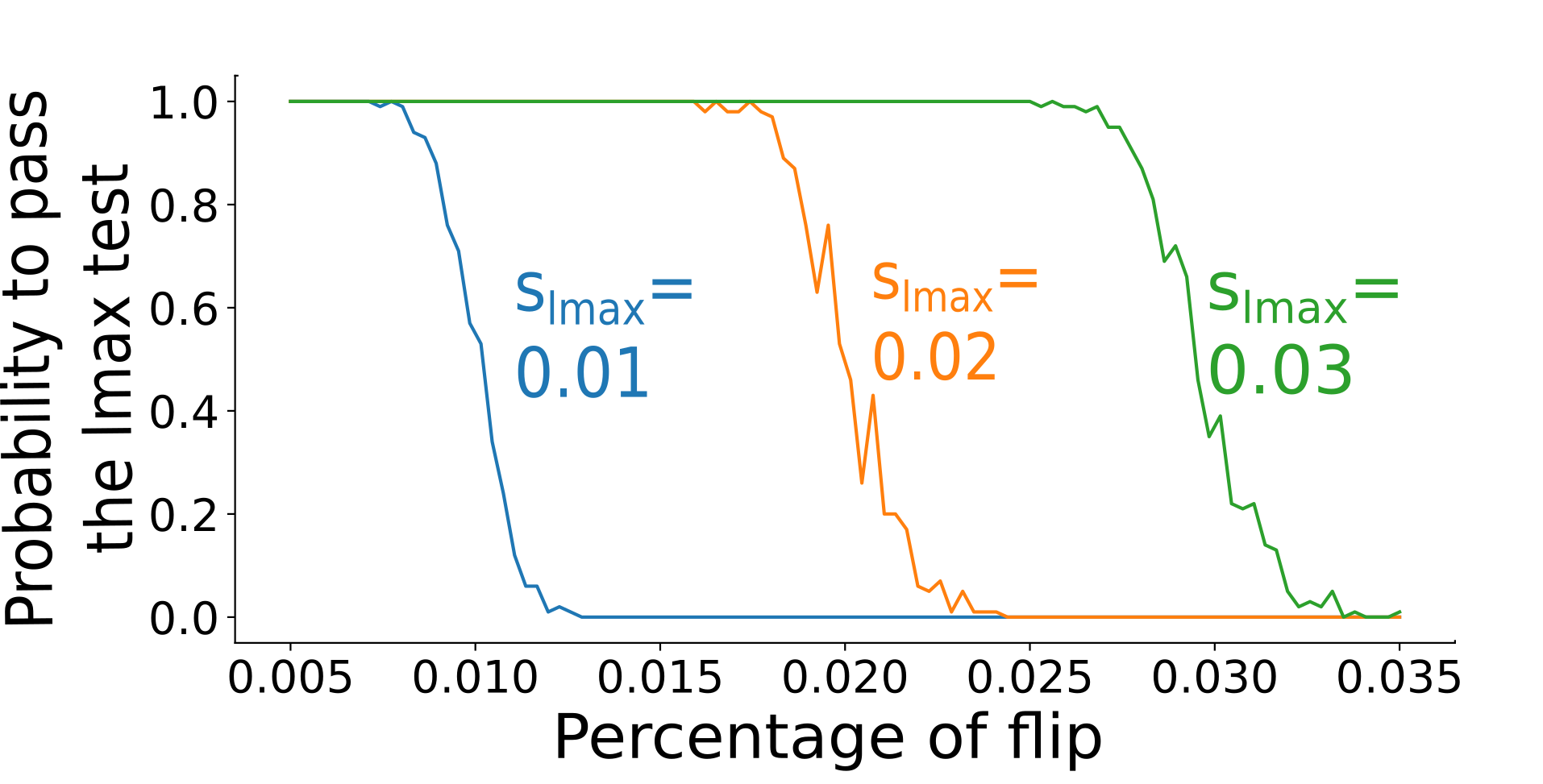}
	\caption{Probability of passing the hardest test (lmax) with different percentage of bitflip applied to the keys.}
	\label{proba_fail}
    \end{center}
\end{figure}

By adding a certain percentage of artificial bitflips at random in our data, we simulated errors in the QKD keys. The errors are added once the keys are generated, more specifically, between the first and second step of the preparation stage on Fig \ref{data_per_users}. Each bit of the hash keys has a chance to be flipped. Fig \ref{proba_fail} shows that changing $s_{l_{max}}$ (the verification condition for the highest transferability level)
changes the resistance of the protocol to errors. The condition for the verification is to have a fraction of errors in the correlations between the tags lower than $s_{l}$.

We see in Fig \ref{proba_fail} that the protocol can be successful even with errors in the data, meaning that the solution of optimising the number of keys even if it adds errors can be an interesting way to make this protocol more efficient for large networks.

We can also notice on Fig \ref{data_error} that the red, green and blue curves show a step at some point. The hash keys shared during the sharing stage have to be identified for the verification stage. Hence, to know which hash key should generate which tag, each hash key has to carry an identification number. The number of keys shared through the network is high enough to make the size of the ID number visible on the curves of Fig \ref{data_error)}. The step we see are these ID numbers going up one bit in size. This ID number can be a source of optimization by putting an upper bound to the number of hash functions shared to avoid having one more bit in the ID number.

\end{document}